\documentclass[11pt]{article}
           \usepackage{color} 

\usepackage{times}
\usepackage{graphicx}
\usepackage{amssymb}
\usepackage{epstopdf}
\begin{document} 

\title{Thermal Electron Wavepacket  Interferometry: Derivation and New Results} 
\author{  E. J. Heller$^{1,3,\ast}$, K.E. Aidala$^{2}$, B.J. LeRoy$^{1}$,  A.C. Bleszynski$^{1}$,   
 A. Kalben$^3$ , R.M. Westervelt$^{1,2}$, \\
 K.D. Maranowski$^{4}$, and A.C. Gossard$^{4}$\\
\\
  $^1$Department of Physics,  
Harvard University, Cambridge,
  Massachusetts 02138 \\
  $^2$Division of Engineering and Applied Sciences,  
Harvard University, Cambridge MA 02138 USA \\
  $^3$Department of Chemistry and Chemical Biology, 
 Harvard University, Cambridge MA 02138 USA\\
 $^4$Materials Department, University of California,  
  Santa Barbara, Santa Barbara CA 93106 USA\\
$^\ast$To whom correspondence should be addressed; E-mail:  heller@physics.harvard.edu  }

  \maketitle

\maketitle

\begin{abstract}

 A novel formal equivalence between thermal averages of coherent  properties (e.g. conductance), and time averages of a single wavepacket arises for Fermi gasses and certain geometries. In the case of one open channel in a quantum point contact (QPC), only one wavepacket history, with wavepacket width equal to thermal length, completely determines thermally  averaged conductance. The formal equivalence moreover  allows  very simple physical interpretations of interference features surviving under thermal averaging.  Simply put, pieces of thermal wavepacket   returning to the QPC along independent paths must arrive at the same time in order to interfere.    Remarkably, one immediate result of this approach is that higher temperature leads to narrower wavepackets and therefore better resolution of events in the time domain.  In effect, experiments at 4.2 K are performing time gated experiments at better than a gigahertz.  Experiments involving  thermally   averaged  ballistic conductance in 2DEGS are presented as an application of this picture.
\end{abstract}

\section{Introduction}

Quantum coherence and control has become a major theme in device physics, with applications in quantum information and computing as possible rewards for success.  The quantum world offers  new phenomena involving interference and entanglement which are not incorporated into present day devices.  The ability to store, process, and retrieve information coherently may lead to qualitatively new technologies.  

Finite temperatures tend to destroy quantum effects, especially if $kT$ is much larger than the mean level spacing of the system.  However thermal averaging is not decoherence; it is 
 not the same as entanglement of the ``system'' with many other degrees of freedom.  For, if we  thermally   average over states which are individually coherent with respect to some degrees of freedom of interest, that coherence remains in some sense.   Well known examples from optics include the ``white light interferometer'', which operates even with blackbody radiation to give high contrast interference fringes. 
 
Coherent wavefunction descriptions of matter waves  give way to incoherent density matrix descriptions at finite temperature. The common wisdom is that a single coherent wavefunction will not suffice when describing a system at thermal equilibrium.  Although this is usually true, there is an exception  which exists in principle for non-degenerate systems.  Consider an operator $ A$ and a system with Hamiltonian $H$ with a non-degenerate set of eigenfunctions.  Suppose a diagonal density matrix $\rho_T$ correctly describes the physical situation:
\begin{equation} 
\rho_T= 
\sum\limits_n p_n\  \vert \psi_n\rangle\langle \psi_n\ \vert
\end{equation} 
thermal  average of the operator $\hat A$ is then $\langle \hat A\rangle _T = {\rm Tr}[\rho_T \hat A]$. 
Now construct a coherent quantum wavefunction
\begin{equation} 
\phi(0) = \sum\limits_n \sqrt{p_n } \ \exp[i \chi_n]\  \vert \psi_n\rangle
\end{equation} 
Then the time average of $\hat A$ over the history of the wavepacket is then 
\begin{equation} 
\label{avg}
<\hat A>  = \lim\limits_{\tau \to \infty}{1\over \tau}  \int\limits_0^\tau \langle \phi(t) \vert \hat A\vert \phi(t) \rangle\ dt = 
\sum\limits_n p_n \langle \psi_n \vert \hat A\vert \psi_n \rangle ={\rm Tr}[\rho_T \hat A ] = \langle \hat A\rangle _T,
\end{equation} 
i.e. the time average of the operator $ \hat A$ over the ``thermal wavepacket'' $\phi(t)$ is the correct density matrix trace of $\hat A$.  The phases $\chi_n$ may be chosen arbitrarily, or chosen so as to control the initial location of the wavepacket. Thus the history of a single, coherent wavepacket determines thermal average. 

There are several  reasons why this is normally not very useful.  First, it is usually not possible to construct {\it a priori} a wavepacket that has the desired  $p_n$'s.  For example, in a closed billiard any compact, analytically simple wavepacket  will have wildly fluctuating  $p_n$'s.  Second, very long time propagations may be needed to converge the time average, a costly and usually non-intuitive process.   Third,  time averaging won't resolve any exact degeneracy and cannot provide a proper incoherent average over the degenerate states. Since degeneracies abound in real systems, these systems are   excluded (this would seem to exclude all scattering systems except possibly 1D wires).  Fourth, even if the  $p_n$'s can be realized in a pre-determined wavepacket, their form  is often such that the wavepacket is not intuitively or computationally useful.  For example, neither the Boltzmann distribution nor the  Fermi distribution gives a compact wavepacket if $k_BT$ is greater than the lowest characteristic energies of the system.  

Thus it is remarkable that all four of these  problems moderate or disappear altogether for at least one very important experimental situation: determination of conductance $G$ of electrons at finite temperature through quantum dots and devices coupled to thermal reservoirs via quantum point contacts (QPC's).  Indeed for a single mode contact we shall show that a nearly Gaussian wavepacket of width equal to thermal length $\ell_T = \nu_F/\pi k_B T$ where $\nu_F$ is the Fermi velocity, can be run to give thermally  averaged conductance and other  thermally   averaged properties.

Thermal averaging is normally thought of as diluting or washing out specific quantum information. Here, we show thermal averaging permits the deconstruction of the  ordering of events in an ersatz time dependent experiment wherein a narrow electron wavepacket is injected into the device. The width of the ersatz wavepacket {\it decreases} with increasing temperature and the implicit time resolution {\it improves}. This raises the  hope that finite temperatures might be harnessed to enhance quantum coherent control.  

 \section{Thermal wavepackets}

 Some time ago it was demonstrated that narrow waveguide channels exhibit quantized condutance, in accordance with the number of energetically allowed transverse modes of the guide \cite{Wees, Wharem}.  Such conductance quantized QPC's  are now standard equipment in the field of two degree of freedom electron gas (2DEG) physics. We will mostly consider the limit in which 0 or 1 channels are open (below or on the first conductance plataeu).  Consider electrons at fixed energy $E$, using the language of scattering theory and the scattering matrix.  Let $\bf t$ be the part of the scattering matrix describing transmission from left to right across the contact. Then, if the lowest transverse mode dominates the transmission (true from the tunneling limit  and onto the  first conductance plateau)  the matrix ${\bf T}(E) \equiv {\bf t}(E){\bf t}(E)^{\dagger}$ has dimension $N$, where $N$ is the number of modes in the left lead.  However ${\bf T}(E)$  has matrix rank one, and has one nonzero eigenvalue $\tau_1(E)$;    the remainder, $N-1$ of them,  are $0$. The channel which conducts with probability $\tau_1(E)$   is a particular linear combination of the usual transverse modes of the leads on the left. This linear combination is given by the eigenvector corresponding to the nonzero  eigenvalue $\tau_1(E) $ of ${\bf T}(E)$.  When ${\bf T}(E)$ is rank one, an arbitrary transverse mode (or incoming plane wave if we choose) impinging from the left produces the {\it same} state on the right as any other incident mode, apart from an overall complex amplitude.

Consider a clean, adiabatic  QPC with one open mode and therefore conductance  $ {2 e^2/ h}$ lying between two perfect reservoirs.  We introduce various ``imperfections'' to the right of the QPC:  scattering  impurities, moveable obstructions (such as the potential hill induced by a charged scanning probe tip\cite{Topinka1,Topinka2}),  charged gates making confined dot regions,   donor potential fluctuations, etc.  These  structures  typically have the effect of reducing the overall conductance below ${2 e^2/h} $, because they cause some scattering back through the QPC.  (However, if they induce resonances on the right side, this can {\it increase} the flow through the QPC, so that the conductance for example reaches $2e^2/h$ even in the nearly pinched off regime.  See refs. \cite{Katine,Blaauboer}.)

 We wish to determine  thermally   averaged conductance, including the structures on the right . 
The conductance is  given at a single energy  by \cite{buttiker,datta}
 \begin{equation}
 G_E = {2 e^2\over h} {\rm Tr} [  {\bf t}(E){\bf t}(E)^\dagger] = {2 e^2\over h} \tau_1(E)  
  \end{equation}
 For an adiabatic, reflectionless contact between two clean, open reservoirs, on the first  conductance plateau, $G_E= {2 e^2/h}$; i.e. $\tau_1(E)  = 1$ in that energy region.
The  thermally  averaged conductance is given by\cite{datta} 
  \begin{equation}
  \label{Gtemp}
 G_T = {2 e^2\over h}\int \ \left ({-\partial f_T(E)/ \partial E}\right ){\rm Tr} [  {\bf t}(E){\bf t}(E)^\dagger]   \ dE  
   \end{equation}
    where $f_T$ is the Fermi distribution function at temperature $T$.
Assuming that the second (excited) and higher transverse modes in the contact are energetically out of reach, the thermally  averaged conductance Eq.~\ref{Gtemp} is an average over  channel $1$ only:
     \begin{equation}
  \label{Gtemp1}
 G_T = {2 e^2\over h}\int \ \left ({-\partial f_T(E)\over \partial E}\right )\tau_1(E) \ dE .
  \end{equation}

We denote by $\psi_1(x,y,E) $ the total scattering eigenstate with the incoming boundary condition of precisely the eigenchannel  of ${\bf t}$ which propagates with with probability $\tau_1(E) $  through the QPC (and which gives the  ${2 e^2/h}\ \ \tau_1(E)  $ conductance, before the structures on the right were introduced).  All the other degenerate channels  orthogonal  to 
$\psi_1(x,y,E) $  have zero transmission.    
 Consider  the wavepacket 
 \begin{equation} 
\label{thermal}
\psi_T(x,y,0)=  \int dE \ a_T(E)\  
e^{i \varphi(E)}\  \psi_1(x,y,E) .
 \end{equation} 
 where $a_T(E) = \hbar^{-1}\sqrt{{-m\ (\partial f_T(E)/ \partial E})/2 \pi }$, which gives a unit normalized wavepacket.
The phase $ \varphi(E)$ can be chosen so that $\psi_T(x,y,0) $ represents a compact  wavepacket  impinging  on the QPC from the left.  Subsequent time evolution then leads to the wavepacket heading through the QPC, with partial reflection immediately from the region of the QPC  if $\tau_1(E) $ is less than unity.  The probability of immediate reflection, due to the QPC itself, is 
  \begin{equation} 
\label{reflect}
R_T^{ immediate}= \left ({2\pi \hbar^2\over m}\right ) \int \ \vert a_T(E)\vert^2 \sigma_1(E)\ dE
 \end{equation} 
 where $ \sigma_1(E)= \sum_l \vert S_{1\ell} (E)\vert^2$ is the reflection probability of the clean QPC from the left, i.e.  $ \sigma_1(E)+   \tau_1(E)=1$.
   The time evolved thermal wavepacket is
  \begin{equation} 
\label{thermal2}
{\psi_T(x,y,t)=\int dE \ a_T(E)\  
e^{-iEt/\hbar+i \varphi(E)}\  \psi_1(x,y,E) }
 \end{equation} 
Far to the  right, $ \psi_1(x,y,E)\to {\bf t}_{11}(E) \psi_1^{out}(x,y,E)$, where $ \psi_1^{out}(x,y,E)$ is a unit flux outgoing wave far to the right, beyond all obstructions.  Then after a time when the wave has vacated the scattering zone, the norm of the wavefunction on the right is easily shown to be
 \begin{equation} 
\label{thermal3}
\left ({4\pi e^2\hbar\over m}\right ){\int dE  \ \vert a_T(E)\vert^2\  
|{\bf t}_{11}(E) |^2 = {2 e^2\over h} \int dE \left ( {-\partial f_T(E)\over \partial E}\right )\  
\tau_1(E) = G_T},
 \end{equation} 
 i.e. {\it the thermally  averaged conductance is obtained by propagating a single  coherent wavepacket from the left and checking how much of it remains on the right.  }
 
 An equivalent  approach is to integrate the flux returning through the QPC over time, or the flux heading to the right.  For incoming channel $\ell$, from the left, the wavepacket is, on the left, 
 \begin{equation} 
\label{wpin}
\psi_T(x,y,t) = \int \ dE \ a_T(E) \ e^{-iEt/\hbar} \left \{e^{i k_\ell(E) x}\psi_\ell(y) +\sum\limits_{\ell'} S_{\ell,\ell'}(E) e^{-i k_{\ell'}(E) x}\psi_{\ell'}(y)  \right \}
 \end{equation} 
On the right, it is (in the asymptotic region beyond obstructions etc.)
 \begin{equation} 
\label{wpin}
\psi_T(x,y,t) = \int \ dE \ a_T(E) \ e^{-iEt/\hbar} \sum\limits_{r'} T_{\ell,r'}(E) e^{i k_{r'}(E) x}\psi_{r'}(y)  
 \end{equation} 
Using the flux operator 
$\hat j = \hbar/m \ \ {\rm Im}  (\psi^*_T\nabla \psi)$,
 in either case we easily 
 recover Eq.~\ref{thermal3} for the conductance.   The time average of the flux operator using thermal wavepacket thus gives the conductance. The flux approach can be extended to calculate the thermally   averaged cross conductance matrix elements  $G_{pq,T}$ in a multiterminal arrangement by averaging the flux out of terminal $q$ given it was sent in terminal $p$. 
 
\begin{figure}
\centerline {
\includegraphics[width=6in]{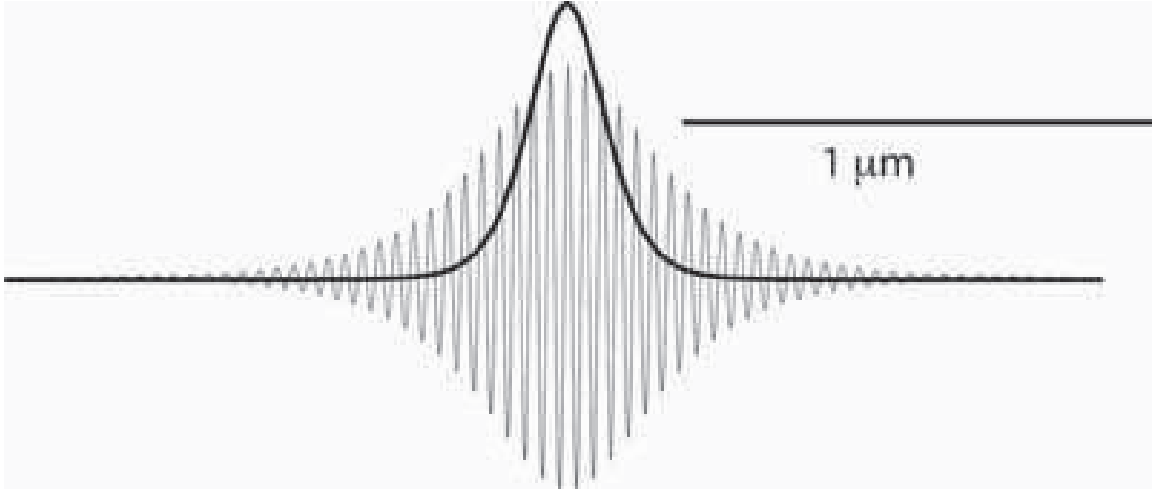}
}
\caption {  The real part and absolute value squared of the  thermal wavepacket at $4.2 $K in a GaAS-AlGaAs heterostructure interface.    }
\label{f1}
\end{figure}

 The approach works because 1) time averaging removes arbitrary phases between states of different energy, 2) there is no degeneracy since by assumption only one mode can pass through the QPC and emerge to the right, and 3) we can construct a simple wavepacket whose amplitude in the eigenstates is exactly 
 
 \begin{equation} 
 w(E) = \sqrt{-\partial f_T(E)/\partial E}={(4 k_B T)}^{-1/2}\  {\rm Sech}[(E-E_F)/2k_B T]
 \end{equation}
  and 4) the amplitude $a_T(E)$ is nearly a Gaussian peaked at $E_F$, (see below) leading to a compact thermal wavepacket.   It should be clear from the derivation that if there are two open transverse modes in the incoming QPC, then we need to run two independent wavepackets, one in the lowest and another in the next transverse mode, etc.  
 We may also compute for example the average wavefunction on the right by time averaging the wavepacket, etc.

 thermal wavepacket idea is especially well motivated for Fermi gasses with an applied bias voltage between leads, where the relevant weighting function $w(E)$
 is approximately  a  Gaussian, peaked at the Fermi energy.  This means thermal wavepacket is itself  nearly Gaussian,  which is {\it more localized the higher the temperature,} and which has a well defined mean velocity and dispersion in velocity.  (Although the Fourier transform giving the thermal wavepacket is not a closed form, there is no requirement  to approximate the wavepacket by a Gaussian; the Gaussian approximation is valuable intuitive purposes but the exact thermal wavepacket may be found numerically if desired). The width of the wavepacket is just the thermal length: $\ell_T = $0.4 microns at 1.7K, and 0.16 at 4.2K in a sample with $E_F = 0.016 eV$ and $\nu_F=2.86 \times 10^5$ m/s. The  thermal length divided by the wavelength is $\ell_F/\lambda_F = E_F/\pi^2 k_B T \sim 5$ in this example.  The momentum uncertainty of this thermal wavepacket, $\delta \nu/\nu_F \sim k_B\ T/2E_F$ is only about 1\% of the average momentum, so that in a 10$\mu$m  round trip the wavepacket gains only about $0.1\mu$m in width at $4.2 $K. Higher temperatures will shorten the thermal wavepacket, and increase its momentum uncertainty. The phase coherence length $\ell_\phi$\cite{Quinn}  provides an upper limit to the  temperature  for practical use of the thermal wavepacket, since the  $\ell_\phi$  needs to be longer than $\ell_T$; this is assured if $k_B T << E_F $, which is normally not a limitation.   Journeys longer than $\ell_\phi$ will be of course not be coherent.  The real part of a thermal wavepacket at  $4.2 $K and $E_F = 0.016 ev$ in GaAs/AlGaAs is shown in Fig.~\ref{f1}.
 
 \section{Physical picture}
 
\begin{figure}
\centerline {
\includegraphics[width=6in]{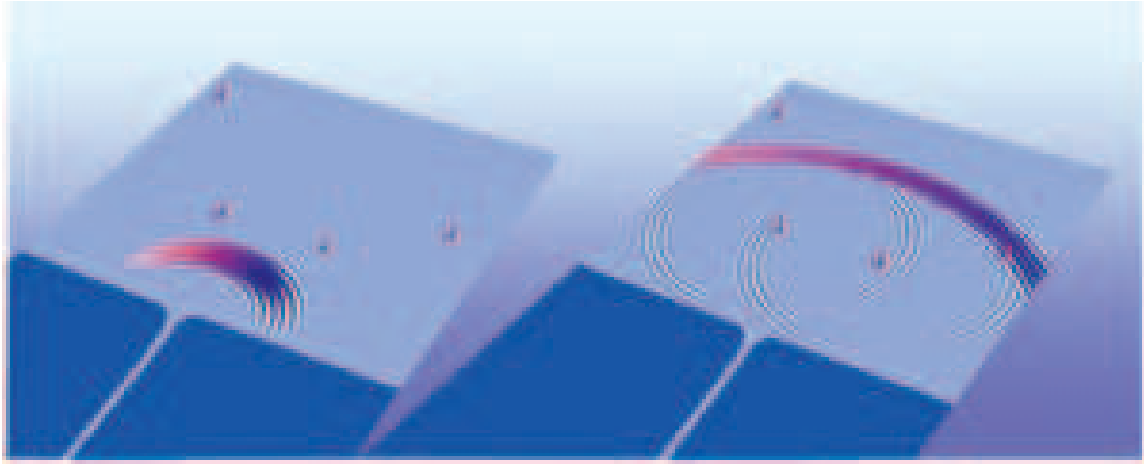}
}
\caption {thermal wavepacket emerging from the QPC encounters two impurities which are approximately the same radius from the QPC; the amplitude from these can interfere at the QPC (right); the amplitude scattered from impurities encountered later cannot interfere with the amplitude from the first two, since the arrival of that amplitude  will be  too late. }
\label{f2}
\end{figure}

 The thermal wavepacket approach permits a powerful intuitive picture  and a simple estimation tool for understanding interference effects and electron choreography in small devices.  The basic scenario is as follows: Launched from just to the left of the QPC, a spatially narrow wavepacket emerges to the right, fanning out as part of an annulus, then suffering small angle scattering,  splitting up into pieces due to collisions with large and small objects, walls, etc. (Fig.~\ref{f2}). The pieces will have the same width in their direction of travel as the original wavepacket (unless the scattering is resonant and therefore time delaying).  As the pieces arrive at the original QPC (or another terminal) the rule is simple: separate pieces of the wavepacket must arrive at the same time  if they are to interfere.  (If the pieces exit to a lead with several modes open, they interfere mode by mode). If an object which the wavepacket encounters is moveable, we may follow the oscillation (in conductance) as the interference of the objects' phase changes from constructive to destructive and back to constructive as it is moved in certain directions.  

We now apply the thermal wavepacket to explain several situations, including recently published and previously  unpublished experimental results.

\subsection{Adiabatic and nonadiabatic leads}

Consider a wave  traveling in a smooth coherent wire, in the lowest transverse mode. Let the profile of the wire open up slowly, as in a trumpet bell, forming one half of a QPC.  Intuition suggests that nearly 100\% of the wave will emerge without reflection. One argument for this is that the wavelength is short compared to the size of the ``bell'' , and if ``slowly'' means small changes over a wavelength there is hardly any reflection.  Calculations bear this out.  (It is interesting that sound waves however have no zero point restrictions; the wavelengths can be  on the order of the length of the trumpet, much larger than the bell. This causes significant nonadiabatic reflection at the bell, especially in the lower registers). The emerging, spreading, unreflected  quantum wave defines a reflectionless {\it incoming}  channel $\psi_0^-$, by time reversal.  An adiabatic QPC with one open mode has transmission unity (assuming adiabatic bells on both ends and a smooth waveguide between them) and $G = 2e^2/h$. 

Suppose now that one moveable obstruction is encountered some distance past the QPC.  We consider an adiabatic QPC and a perfect ballistic conductor with no backscattering other than the moveable obstruction. The wave returning from this obstruction will have some component in the $\psi_0^-$ channel, and this will transmit back through the lead without reflection. Other channels will be 100\% reflected with no transmission.  Neglecting for a moment any double or multiple bouncing off the obstruction, no other returning flux exists, and thus {\it no interference pattern will result as the obstruction is moved away from the QPC}, even at very low or 0  temperature.  Since the neglect of multiple bouncing is a good approximation at long distances, the prediction is that the fringes will disappear at larger radii, even at very low temperature.   



\subsection{Fringing}

In a previous publication\cite{Topinka1},  conductance fluctuations  were reported in a QPC device  as a charged  scanning probe microscope (SPM) tip was scanned across the sample at long range from the QPC. The system was  free of obstructions on both sides of the QPC (except for unavoidable impurities and donor atom potential flctuations) There were three striking aspects of this data.  First was its high spatial resolution, which was  explained as due to the ``glint'' effect from SPM tip divot, much smaller than the divot itself\cite{PT}.  Second, the flow was found to be strongly branched, a phenomenon which arises from a concatenation of caustic structures from the small angle deflections caused by donor atom potential fluctuations\cite{Topinka1,Shaw}. Third, the electron flow branches were decorated with interference fringes, even very far from the QPC. These fringes are robust to thermal averaging, which at first is surprising.  For example, if the fringing was due to interference from  imperfections near the QPC and the backscattering from the tip, they should have died out in well under a micron from the QPC, due to averaging over different wavelengths in the thermal average.

\begin{figure}
\centerline {
\includegraphics[width=5in]{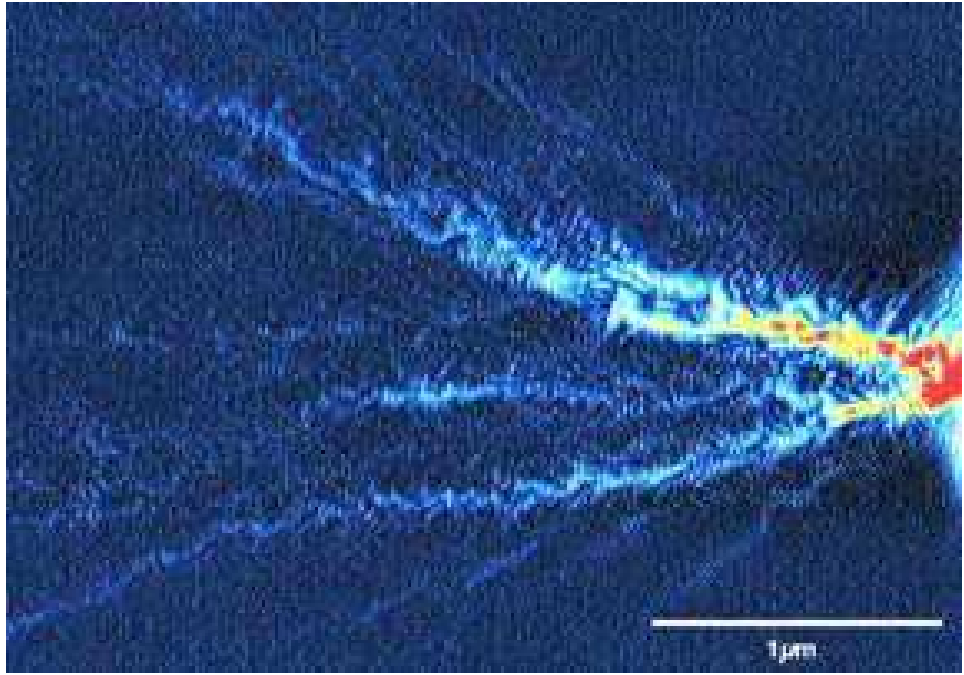}
}
\caption {Branching and quantum interference fringing of electron flow as imaged by the charged SPM technique (Ref.\cite{Topinka1}). }
\label{f0}
\end{figure}

We can easily explain the persistent fringing using the thermal wavepacket picture.  As the electron wavepacket emerges from the QPC, it encounters the small impurities, which each scatter an s-wave in all directions as the outgoing wave passes by. (See Fig,~\ref{f2}). Some of this amplitude is directed back towards the QPC.  We assume (as is the case in the experiments) that the impurities are sparse enough that multiple scattering from the impurities is a negligible contribution at the QPC. The SPM tip also scatters the wave coming from the QPC in many directions, although not uniformly, as it is a larger target.  Some of this amplitude also heads back toward the QPC.  (If the SPM ``divot'' is several wavelengths across, then the scattering in a given direction can be attributed to just a part of the surface of the divot- the ``glint'' effect.)
As the SPM amplitude returns to the QPC, any amplitude  which is also returning after scattering off impurities can interfere.  It is thus immediately clear that only those impurities within a half a thermal length of the SPM tip are able to effect fringing. The net impurity amplitude arriving at the same time as the SPM tip amplitude is a fixed background against which the tip amplitude beats as the tip is moved radially away from the QPC.  Since the thermal wavepacket is typically many wavelengths wide in its direction of travel, the main effect of moving the tip is to cause a phase shift of its amplitude and thus regular fringes in the conductance.  Larger radial shifts of the SPM (on the order of the thermal length) mean that the SPM amplitude is interfering with amplitude from a new set of impurities. This picture explains why the fringes seen in the experiment ``aim'' back to the QPC (i.e. the normal to the fringe antinodes points back to the QPC). Constant radius movement of the tip does not change the phase of its returning amplitude at the QPC, and thus no fringes are crossed in this direction. According to this picture, there should be no die off of the fringing at large radius, for distances much less than $\ell_\phi$, except for the geometric fall off of the SPM tip signal, which decays as $1/r^2$, where $r$ is the radius from the QPC. However the average signal also dies off at this rate, so the fringes remain unabated in relative strength at large $r$, which is in fact observed.  The dynamics of   small angle donor potential scattering leading to branching leads to interesting  corrections to this picture, which can also be explained by the thermal wavepacket model.

  Impurities whose main contribution is to backscatter the direct amplitude coming from the QPC must therefore reside in an annulus of mean radius $R$, of width of the order of the ballistic thermal length in order to contribute to the interference with the tip at radius $R$ from the QPC.  This explains the fringing in the first generation of experiments\cite{Topinka1},  where weak impurity backscattering interferes with the SPM amplitude to cause fringing. 
  
\begin{figure}
\centerline {
\includegraphics[width=4in]{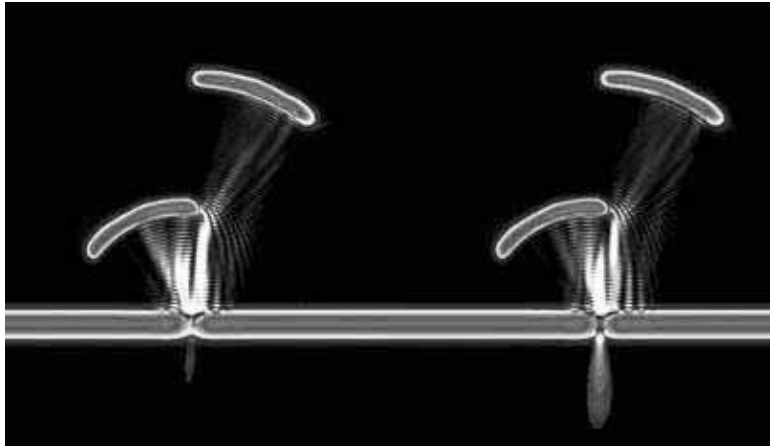}
}
\caption { (left) Snapshot of a thermal wavepacket returning from one bounce off the distant mirror, and two bounces off the close mirror.  A node has developed from the interference between the two, seen as a vertical black region above the center of the QPC,  This wave has difficulty propagating through the QPC, since the excited transverse mode is reflected (close channel).  (right)  This images is taken at the identical moment as the one on the left, with no changes except the outer mirror has been moved about 1/4 of a wavelength.  Now an antinode develops above the QPC, and much more flow back through the QPC occurs, corresponding to lowered conductance. }
\label{f4}
\end{figure}

  Stronger scatterers such as a concave mirror, however, can reflect considerable amplitude back to the walls containing the QPC, from which it can bounce back to be reflected again.  Such a path, which has a length twice that of the single bounce, will only be able to interfere with the direct backscattering from the SPM tip if it is positioned approximately twice as far away as the mirror, so its direct reflecting path has the same length as the double bounce.   At this distance, the single and triple mirror bounce interference arrives too early or too late to interfere with the first bounce from the more distant mirror (Fig.~\ref{f4}) , and the double bounce fringing emerges essentially uniquely at this distance.  This was observed in experiments in which the second mirror was the SPM divot\cite{ScienceNew}.

 Figure \ref{f4} shows the square of a thermal wavepacket which has bounced off two mirrors, in two different cases. In both cases the  wavepacket had earlier emerged from the QPC and was then, roughly speaking,  cut into two pieces by the first mirror. One piece, which hit that mirror, returned toward the wall, and the other piece continued its journey toward the second mirror.  The first piece, after bouncing off the wall and hitting the first mirror again, returns toward the QPC, arriving at the same time as the first bounce from the second mirror.  This is the time at which this figure shows the wavefunctions. They interfere with essentially the classic two slit interference pattern, corresponding to their different paths, which results in a sequence of nodal lines approximately in the direction of travel.  The flux back through the QPC is strongly affected by the alignment of the nodal lines with respect to the QPC.  The nodal lines remain essentially fixed during the arrival of the two packets.  If an independent run is made with either of the two mirrors slightly shifted, the nodal lines shift.  The two cases shown differ by a shift of the outer mirror by about 1/4 of the mean deBroglie wavelength.  Because the wavepacket interacts twice with the close mirror, the rate of shift of the fringes is twice that of the outer mirror.

Another scenario obtains when the divot is close to the QPC:  The amplitude from the first bounce off the divot interferes with paths which have bounced twice off the divot;  this effect was held responsible for the  fringing seen in imaging of  emerging transverse modes near a QPC\cite{Topinka2}.  The first return bounce creates an amplitude against which the double bounce can interfere; since the double bounce aquires a phase relative to the single bounce at the same rate the single bounce does to any fixed amplitude, the double bounce fringes look ``normal'', with a fringe spacing of half a Fermi wavelength.  
However, the double bounce scenario is wiped out by thermal averaging if the total pathlength to the obstruction and back is greater than a thermal length $\ell_T$, about 0.4 microns in \cite{Topinka2}. The thermal wavepacket approach shows why: the single bounce amplitude will arrive back at the QPC and clear away before the double bounce begins to arrive  if the obstruction is more than a thermal length away. The single and double bounces  will then be unable to interfere

Now suppose  the QPC has internal ``defects'', or the energy is such that the electron has to tunnel through it (no open channels), making it nonadiabatic.  The reflection from the divot  into scattering channel $\psi_0^-$ can now travel back through the lead (with partial reflection), emerging on the left to interfere with the amplitude that was reflected from the QPC as it approached from the left side.  If the obstruction is far enough away that the initially reflected amplitude on the left escapes the QPC region to the left  before the backscattered amplitude from the obstruction arrives, then once again there can be no interference between them and no fringing will result.  Suppose this is the case at some temperature $T$.  Then, at some lower temperature, fringing will appear due to this scenario, as the thermal wavepacket becomes more extended.

\section{Quantum control with electron phase}


The effective time resolution $\tau_T$  in the thermal wavepacket (the time it takes to pass a given position) changes with temperature, is $\ell_T/\nu_F = \hbar/\pi k_B T$.  In the familiar case of 4.2 K at $E_F$ = 16 meV, this is already half a picosecond; at 20 K,   $\tau_T = 10^{-13} s$.  Events separated by times greater than  $\tau_T$  behave classically, while quantum interference remains for events within $\tau_T$ of eachother.  This permits filtering of interference ``noise'' as distinct from the ``signal'', by arranging the signal to arrive from separate paths with $\tau_T$. 
 
  \begin{figure}
\centerline {
\includegraphics[width=6in]{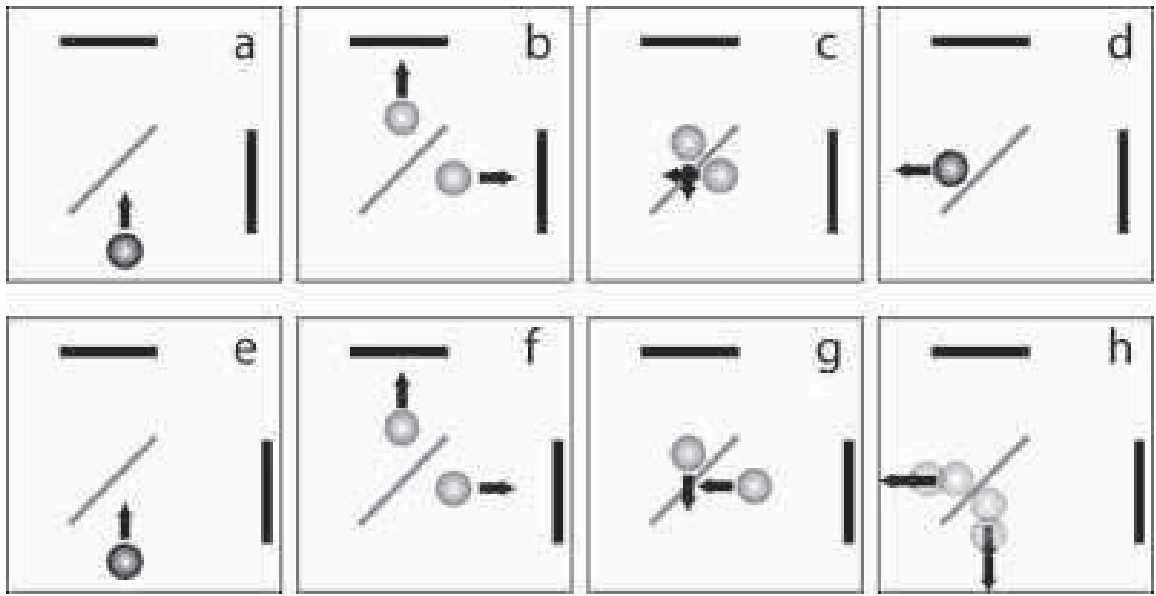}
}
\caption {  A schematic of a feasible Michelson interferometer experiment with electrons. The thermal wavepacket is shown impinging on the 50\% mirror in a. In b, the split packets are heading for their respective mirrors, and in c, they are about to interfere coherently after experiencing equal length round trips  to give d, the constructive interference for the ``left'' as opposed to ``down'' paths.  In e-h, the right mirror has been moved out, so that the wavepackets return at different times and are unable to interfere, spliting into four pieces with no ``left-down'' selection.  The thermal length (divided by 2) determines the minimum distance the right mirror needs to move to remove the interference.   }
\label{michelson}
\end{figure}

A simple realizable  experiment exploits this time resolution, illustrates the power of the thermal wavepacket  viewpoint, and points the way to a new coherent electron switch.  Suppose we construct a Michelson interferometer  involving a 50\% splitter mirror and two 100\% reflection mirrors in the canonical arrangement (Fig.~\ref{michelson}).  This scenario illustrates a coherent control switch for the electron pathway.  When the outer 100\% reflecting mirrors are at nearly equal distance, the contrast is nearly 100\% for left versus down control; the contrast lessens until when one mirror has been moved a thermal length (divided by 2), after which no left-down contrast remains.  The key notion is that both wavepackets returning toward the splitter  must arrive at the same time if they are to interfere most  effectively.  Delays in arrival of one compared to the other mean that the leading edge of one wavepacket, and the trailing edge of the other have no ``partner'' with which to interfere.  Enough movement of one mirror to cause the arrivals at the splitter to be nonoverlapping result in no selection for left versus down, and four wavepackets being produced.

Assuming that the mirrors are equidistant, the Michelson scenario points to the possibility of a ``warm'' but still phase controlled, coherent,  electron switch.  A gate lying above one of  the two paths taken by the electron wavepacket after scattering off the  splitter could phase shift one of the split packets relative to  the other, using a small charge and  with almost no change in arrival time.  This would change the electron path from down to left, for example, after the packets recombine at the splitter.

\section{Simulating higher temperatures}

To demonstrate the validity of the thermal wavepacket picture that has been presented, we have imaged the electron flow from a QPC using a low-temperature SPM in a configuration similar to the two mirror geometry discussed previously.  We lithographically defined a metal gate that acts as one of the mirrors, and our movable tip acts as the other.  Figure 6a shows a schematic of the device geometry and the area of the scan, and subsequent panels show the tip scan data at increasing ``temperature'' (see below).   Interference fringes appear in the image from distinct round trip trajectories of the same length, to within a thermal length.  Increasing the electron temperature decreases the thermal length, and localizes the extent of the fringes, Fig.(\ref{fringedecay}). Note the corresponding size of the thermal wavepacket, which shrinks with increasing temperature. 

To increase the electron ``temperature", we used an arbitrary function generator to generate a time varying voltage signal across the QPC.  The voltage from the function generator injects electrons of that energy through the QPC, and mimics the energy distribution of electrons at the intended temperature. The amount of time at any given voltage is given by the probability of the electron having that energy at the desired temperature. Therefore, the voltage spends the most amount of time at 0 V, which corresponds to the Fermi energy and the smallest amount of time at the highest and lowest voltages, which are the tails of the distribution. As the temperature is increased the voltage spends more time away from 0 V because there is a larger spread in energy of the electrons.

   \begin{figure}
\centerline {
\includegraphics[width=5in]{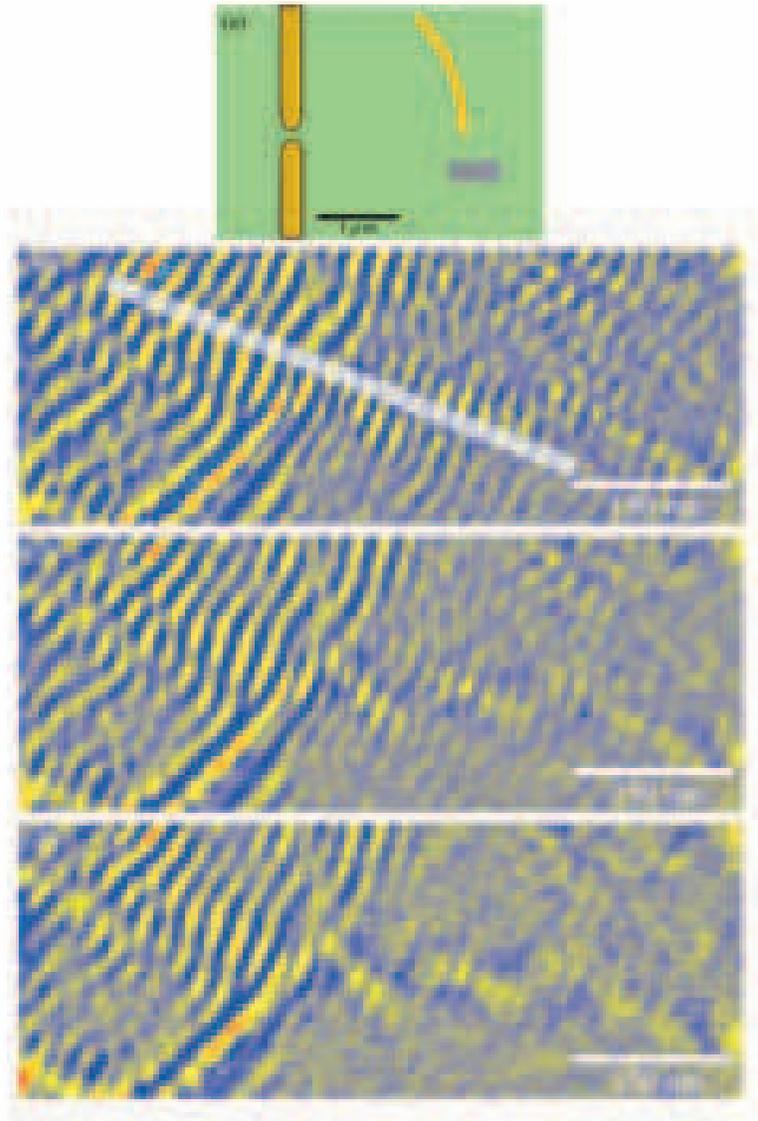}
}
\caption {  (a) Schematic diagram showing the location of the images of electron flow. Images of electron flow at three temperatures, (b) 1.7 K, (c) 4.2 K and (d) 8.4 K showing the robustness of the fringes near the radius of the arc. See the text for an explanation of the procedure used to simulate these temperatues. The white line in (b) shows the location of the one-dimensional line sample used in Figure \ref{fringedecay}.}
\label{strip}
\end{figure}

   \begin{figure}
\centerline {
\includegraphics[width=5in]{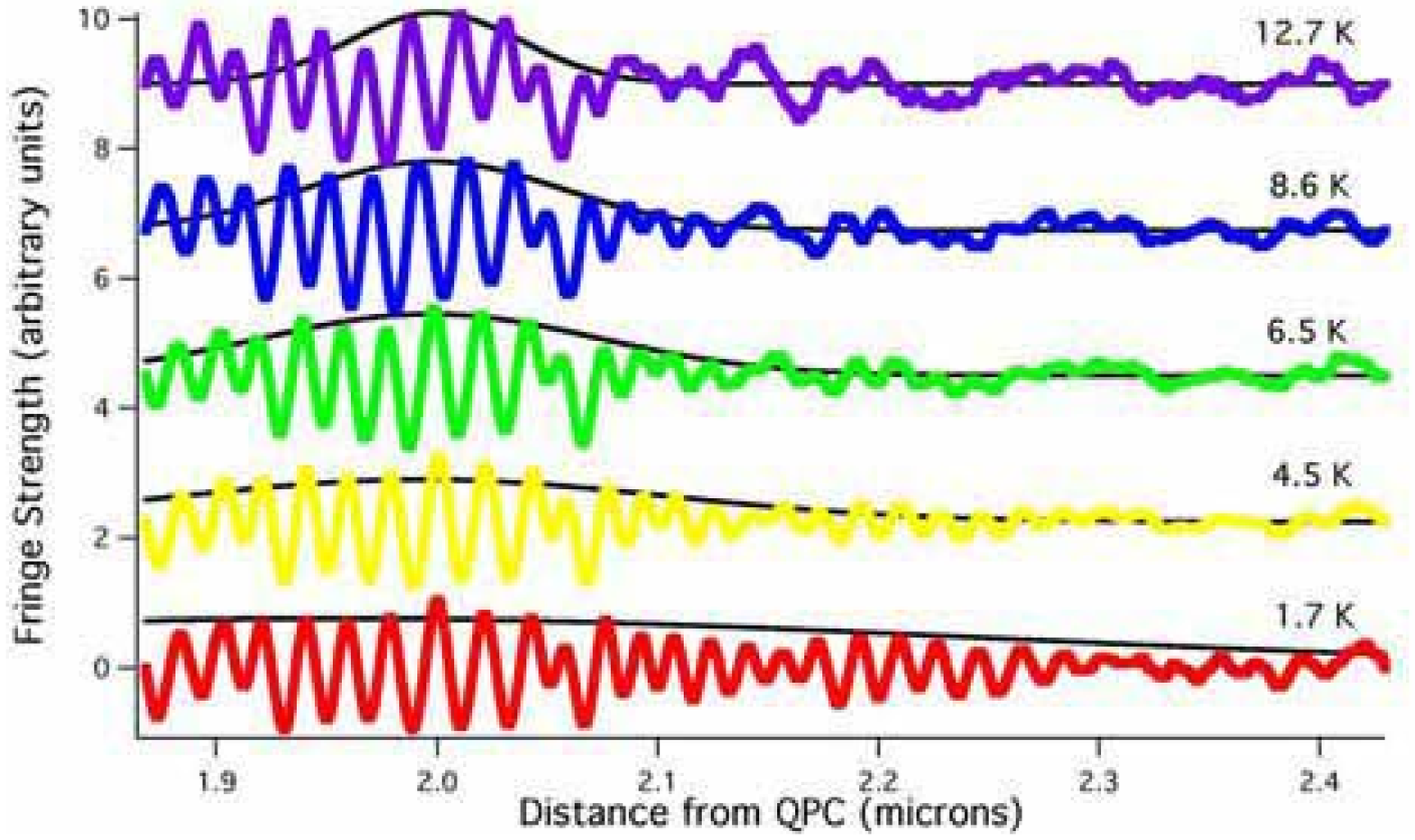}
}
\caption {  Graph of the measured signal as a function of the distance from the arc for a series of simulated temperatures, along the strip shown in Fig.~\ref{strip}. The strength of the signal decays more quickly away from the arc for increased temperatures but remains robust at the arc position.  In each case, a black outline shows the dimension of the corresponding thermal wavepacket}
\label{fringedecay}
\end{figure}

\section{Conclusion}

Thermal electron interferometry  has strong analogies to while light interferometry.  The simplest examples are two slit interference and weak localization coherent backscattering.  The central peak in the two slit case, and the weak localization phenomenon are examples of interference phenomena which survive averaging over different wavelengths.
We have introduced a new way of calculating and understanding  thermally   averaged electron  interference  in semiconductor heterostructures.  A single ``thermal wavepacket''  whose width in the direction of travel is the thermal length can yield thermal averages of measurable quantities such as conductance. Various experiments  were described in terms of thermal wavepacket dynamics, including some earlier data, and some reported here for the first time.    A proposal for a coherent Michelson switch was also given.   

Computing  thermal wavepackets can be an advantageous way to obtain thermal averages numerically. The alternative is to obtain fixed energy solutions and average over them explicitly. Although we do not anticipate here what  methods might be employed to make the fixed energy calculations efficient, the efficiency of fast Fourier transform propagation for  wavepacket dynamics is legion\cite{kosloff}.  Obtaining a theoretical SPM image, which includes all the aspects implied by the presence of the divot,  is a time consuming task, requiring a complete scattering calculation and thermal average for every new position of the SPM tip.  This  gives one pixel of information for each tip position.  Even a modest image of 1000 pixels requires 1000 independent,  thermally   averaged runs.  the thermal wavepacket approach enjoys many advantages in this situation, one of which is re-use of the calculations  up to the time when thermal wavepacket hits the divot.

{\it Acknowlegement:}  We thank Jacob J. Kritch for a critical reading of this manuscript.  This work was supported at Harvard University by DARPA grant DAAD19-01-1-
0659 and by the Nanoscale Science and Engineering Center (NSEC) under NSF grant PHY-0117795;
and NSF CHE-0073544; work at UC Santa Barbara was supported by the Institute for Quantum Engineering, Science and Technology (iQUEST).

\end{document}